# Insulator to Metal Transition under High Pressure in FeNb$_3$Se$_{10}$

Haozhe Wang[1,2], Shuyuan Huyan[3,4], Eoghan Downey[5], Yang Wang[5], Shane Smolenski[5], Du Li[6], Li Yang[6], Aaron Bostwick[7], Chris Jozwiak[7], Eli Rotenberg[7], Sergey L. Bud'ko[3,4], Paul C. Canfield[3,4], R.J. Cava[2], Na Hyun Jo[5*], Weiwei Xie[1*]

1. Department of Chemistry, Michigan State University, East Lansing, MI 48824, USA
2. Department of Chemistry, Princeton University, Princeton, NJ 08540, USA
3. Ames National Laboratory, Iowa State University, Ames, IA 50011, USA
4. Department of Physics and Astronomy, Iowa State University, Ames, IA 50011, USA
5. Department of Physics, University of Michigan, Ann Arbor, MI 48109, USA
6. Department of Physics, Washington University in St. Louis, St. Louis, MO 63130, USA
7. Advanced Light Source, Lawrence Berkeley National Laboratory, Berkeley, CA 94720, USA

*Abstract*

Non-magnetic FeNb$_3$Se$_{10}$ has been demonstrated to be an insulator at ambient pressure through both theoretical calculations and experimental measurements and it does not host topological surface states. Here we show that on the application of pressure, FeNb$_3$Se$_{10}$ transitions to a metallic state at around 3.0 GPa. With a further increase in pressure, its resistivity becomes independent of both temperature and pressure. Its crystal structure is maintained to at least 4.4 GPa.

Corresponding authors: Weiwei Xie (xieweiwe@msu.edu), Na Hyun Jo (nhjo@umich.edu).

## Briefly, on Frank DiSalvo

When the senior author of this manuscript (Cava) was a much younger scientist beginning his career at Bell Labs, he spent countless hours in Frank's Lab near the bathroom on the first floor of Building 1, now demolished, learning about the synthesis of chalcogenides from Frank and his technician, Joe Waszczak, because as any chemist can tell you it's the tricks of the trade, not written explicitly in papers, that lead to high quality syntheses. Frank and Joe were very generous with their time and that was back when the world was a much smaller place. Cava remembers working on $FeNb_3Se_{10}$, the blue bronze, grown electrochemically by Lynn Schneemeyer, and $NbSe_3$ on the magnetometer in Frank's Lab. As we all know, Alzheimer's took Frank away from us mentally years ago, which no doubt tortured his wife, Barbara. As we age, we know that something will eventually kill us, but our loss of Frank, Robert Fleming, Maurice Rice, Ward Plummer, and Daniel Khomskii in the past year has nonetheless not been easy to bear. There but for fortune go us all, impossible as it may be for a young person to comprehend.

Frank and Joe grew exquisite crystals of the dichalcogenides and were connected to the excellent French school of solid-state chemistry through Frank's mentor Mike Sienko. The project described here on $FeNb_3Se_{10}$ has been in the back of Cava's mind for three decades, because he remembers that his friend and lab-next-door neighbor Robert Fleming (also deceased) back in the mid-1990s pressurized single crystals of $FeNb_3Se_{10}$ to 1 GPa and saw that their resistivity decreased. Fortunately, his ex-postdoc Weiwei Xie sometimes listens when Cava mumbles and, in collaboration with the group of Paul Canfield, tested the resistivity and crystal structure of that material to much higher pressure. This study describes the results of high-pressure measurements, combined with ambient pressure ARPES (done by Na Hyun Jo, another alumna from Paul Canfield's group), on $FeNb_3Se_{10}$.

**Introduction to this work**

Transition metal dichalcogenides, when intercalated by magnetically active elements, can often display important magnetic behavior, and can potentially be used for spintronic and other information technology applications.[1,2,3,4] Much research has been conducted on the $4d$ and $5d$ early transition metal dichalcogenides, such as $3d$ element intercalated $Fe_xNbSe_2$, which may even host helimagnetic or spiral magnetic properties.[5] The more Se-rich trichalcogenides such as $TaSe_3$ and $NbSe_3$ have also garnered significant research interest due to their exotic electrical transport properties.[6,7,8,9,10,11,12] Notably, $NbSe_3$ has been extensively studied due to the observation of two incommensurate charge density waves (CDWs) and superconductivity.[12,13,14,15,16,17] The fundamental structural units of $NbSe_3$ are triangular face-shared $NbSe_6$ trigonal prisms, which form three different infinite chains that run parallel to its monoclinic $b$-axis. Substantial cross-linking occurs between chains through Nb-Se bonds, which are only slightly longer than those within the trigonal prisms. The resulting two-dimensional slabs are weakly bonded to neighboring slabs through Se-Se van der Waals interactions.

However, when incorporating some $3d$ elements into $NbSe_3$, intercalation between the two-dimensional $NbSe_3$ layers does not happen, instead, the $3d$ ions mix with Nb atoms and form a new type of compound, which we write here as $TNb_3Se_{10}$ (T= Cr, Fe and Co) although various studies, including this one, indicate that the materials are $3d$ transition-metal rich.[18,19] The crystal structure of $FeNb_3Se_{10}$ consists of two $NbSe_6$ trigonal prismatic chains and a double chain of edge-shared $(Fe,Nb)Se_6$ octahedra. Previous studies showed that the resistivity of $FeNb_3Se_{10}$ increases by nine orders of magnitude on cooling from 140 K to 3 K, in contrast to $NbSe_3$, which exhibits metallic behavior under similar conditions.[20,21,22,23] X-ray scattering revealed that like in $NbSe_3$ the CDW in $FeNb_3Se_{10}$ is incommensurate, with a wave vector $q$ of (0.0, 0.27, 0.0); along the chain axis as is the case for $NbSe_3$.[24,25]

In order to better understand $FeNb_3Se_{10}$, we studied its electrical properties at both ambient pressure and high pressure. We confirmed that $FeNb_3Se_{10}$ is an insulator at ambient pressure through both theoretical calculations and experimental measurements. Upon the application of pressure, $FeNb_3Se_{10}$ transitions to a metallic state around 3 GPa and with further increase in pressure, it becomes a bad metal at around 7 GPa. Although $FeNb_3Se_{10}$ is structurally related to

NbSe$_3$, bulk superconductivity has not been observed in FeNb$_3$Se$_{10}$ at temperatures above 1.8 K, consistent with the behavior of bulk NbSe$_3$..

## Experimental and Computational Details

**Single Crystal Growth:** Single crystals of FeNb$_3$Se$_{10}$ were synthesized via chemical vapor transport.[26] The high purity elements Fe, Nb, and Se were ground into powder, homogenously mixed in the stoichiometric ratio of 1:3:10, and then heated in sealed evaluated quartz tubes under a vacuum of less than 10$^{-5}$ torr. The quartz tubes were heated to 700 °C in a single-zone tube furnace and maintained at this temperature for one week. The two ends of the furnace reaction tube were packed with quartz wool to minimize the temperature gradient. Subsequently, the tubes were cooled to room temperature at a controlled rate of 10 °C per hour. Needle-shaped crystals (around 1 mm long ribbon-like crystals) were collected from the end of the quartz tube opposite to the basic charge.

**Single Crystal X-ray Diffraction (XRD) at Ambient Pressure:** A single crystal with the dimensions of 0.561 × 0.032 × 0.021 mm$^3$ was picked up, mounted on a nylon loop with paratone oil, and measured using a XtalLAB Synergy, Dualflex, Hypix single crystal X-ray diffractometer with an Oxford Cryostreams low-temperature device, operating at room temperature. Data were measured using $\omega$ scans using Mo K$_\alpha$ radiation ($\lambda$ = 0.71073 Å, microfocus sealed X-ray tube, 50 kV, 1 mA). The total number of runs and images was based on the strategy calculation from the program CrysAlisPro 1.171.43.92a (Rigaku OD, 2023). Data reduction was performed with correction for Lorentz polarization. An empirical numerical absorption correction based on Gaussian integration over a multifaceted crystal model was applied. The structure was solved and refined using the Bruker SHELXTL Software Package.[27]

**High Pressure Single Crystal XRD:** To investigate the effects of pressure on the crystal structure, a high-pressure single crystal XRD experiment was performed on a single crystal of FeNb$_3$Se$_{10}$ up to 4.4 GPa. Prior to the high-pressure experiment, the sample was mounted on a nylon loop with paratone oil and measured at ambient pressure to determine its crystal structure. The sample was then loaded into the Diacell One20DAC manufactured by Almax-easyLab with 500 μm culet-size extra aperture anvils. A 250 μm thick stainless-steel gasket was pre-indented to 76 μm and a hole of 210 μm was drilled using an electronic discharge machining system. A 4:1 methanol–ethanol mixture was used as pressure transmitting medium.[28] The pressure in the cell was monitored by the R1 fluorescence line of ruby.[29]

**Angle Resolved Photoemission Spectroscopy (ARPES):** ARPES experiments were conducted at Beamline 7.0.2 (MAESTRO) at the Advanced Light Source. The data were acquired using the micro-ARPES end station, which consists of an Omicron Scienta R4000 electron analyzer.

Samples were cleaved in situ by carefully knocking off an alumina post (0.28 mm in diameter) attached to the top of each sample with silver epoxy. Due to van der Waals bonding, samples were cleaved well along the crystallographic *c*-axis. Data were collected with photon energies of 138 eV. The beam size was ~ 15 μm × 15 μm. ARPES measurements were performed at $T$ = 27 K under a vacuum (UHV) better than $4 \times 10^{-11}$ torr.

**Electrical Transport Measurement:** Resistance measurements down to 1.8 K were conducted using a Quantum Design Physical Property Measurement System (PPMS) Dynacool using the electrical transport option. To have good electrical contacts, the samples were manually masked into a standard four-probe geometry, and Au films were evaporated to form ohmic contacts. 25-micron-diameter Pt wires were attached to the samples using DuPont silver paint (4929N) in standard 4-probe configurations.

**High Pressure Electrical Resistance Measurement:** Electrical resistance measurements under pressure were conducted using a Quantum Design Physical Property Measurement System (PPMS). A diamond anvil cell (DAC) from Bjscistar was used to fit into the PPMS chamber.[30] The DAC uses diamonds (standard-cut type Ia) with cullet sizes of 700 μm and 300 μm to cover different pressure ranges; 1–8.4 GPa for sample 1 and 1.8–65.5 GPa for sample 2. For the measurements, thin, ribbon-like $FeNb_3Se_{10}$ single crystals were cut into flakes of approximately 200×40×10 μm for the 700 μm culet size and 50×50×10 μm for the 300 μm culet size. The flakes, along with a tiny ruby sphere (< 30 μm in diameter), were then loaded into the sample chamber with a stainless-steel gasket that was 250 μm thick, apertured, and covered with cubic-boron nitride (cBN). Platinum foil served as the electrodes. For sample 1, resistance was measured using a standard four-terminal configuration with the electrical current down the long axis of the ribbon (the [010] direction), consistent with ambient pressure measurements shown in **Fig. 2**. For sample 2, in order to determine resistivity anisotropy, the van der Pauw method was employed with Nujol mineral oil was used as the pressure-transmitting medium (PTM) which prevents direct contact between the sample and the diamond culet, which would otherwise introduce an additional uniaxial pressure component.[31,32,33,34] Pressure was determined by monitoring the $R_1$ line of the ruby florescent spectra.[35]

**Computational Details:** We applied density functional theory with Hubbard U (DFT+U) calculations to obtain the ground state properties of $FeNb_3Se_{10}$. The core-electron states of Iron, Niobium and Selenium atoms were eliminated by the projector augmented-wave (PAW)[36] potentials as implemented in the Vienna ab initio simulation package (VASP)[37]. The Generalized Gradient Approximation (GGA) functional is used in the calculation, and spin-orbital coupling is included.[38] The plane-wave energy cutoff is set to be 400 eV, and a $2 \times 6 \times 2$ $k$-grid is adopted. We choose the Hubbard parameters U = 2.2 eV for the Iron atom, and U = 1.0 eV for the Niobium atoms, respectively to align with the experimental ARPES band measurements.[39]

## Results and Discussion

**The Quasi-one-dimensional Crystal Structure of FeNb$_3$Se$_{10}$:** The crystal structure was determined by single crystal X-ray diffraction measurements. The results show that FeNb$_3$Se$_{10}$ crystalizes in the previously reported monoclinic structure with the space group of P2$_1$/*m*, with five independent Se and two independent metal positions.[18] The metal atoms were found to be in both trigonal prismatic and octahedral coordination to selenium. The refinements show that Fe and Nb both occupy the octahedral sites, in a disordered fashion, in a ratio that suggests that the formula of the phase is iron rich of FeNb$_3$Se$_{10}$, approximately Fe$_{1.25}$Nb$_{2.75}$Se$_{10}$, close to but not exactly that seen in the powder study. Both atomic mixtures and vacancies were refined systematically, and only the octahedra were found to be Nb deficient, with that site having about a 40% mixture of the Nb with Fe atoms. From a charge balance perspective, the formula FeNb$_3$Se$_{10}$ can be considered as (Fe$^{2+}$)(Nb$^{5+}$)$_3$[(Se-Se)$^{2-}$]$_2$(Se$^{2-}$)$_6$, yielding a net charge of +1. Although this indicates that the compound is not entirely charge-balanced, the significant Fe occupancy at the Nb/Fe site may result in a total charge close to zero, thereby approaching charge neutrality. **Fig. 1** shows the crystal structure viewed down the monoclinic *b*-axis. The NbSe$_6$ prismatic chains run parallel to the *b*-axis while the (Fe,Nb)Se$_6$ octahedra are located between two NbSe$_6$ prisms. Each (Fe,Nb)Se$_6$ octahedron shares four edges with neighboring octahedra and one corner Se atom with the NbSe$_6$ trigonal prisms. The octahedrally coordinated Nb and Fe atoms form a double chain parallel to the *b*-axis. The interpolyhedral metal-selenium bonding forms two-dimensional slabs parallel to the *ac*-plane. Similar to NbSe$_3$, these two-dimensional slabs are held together by van der Waals bonds between Se atoms. The Se-Se distances across the van der Waals gaps are comparable to those in NbSe$_3$. This crystal structure was maintained under at least up to 4.4 GPa (highest pressure in our xrd measurements), with the refinement data shown in **Tables 1&2**.

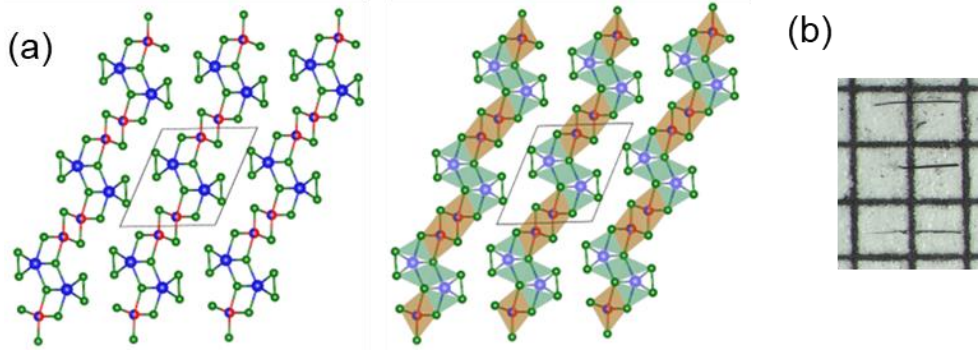

**Fig. 1.** (*a*) The crystal structure of FeNb$_3$Se$_{10}$ projected into the *ac*-plane in two types of representations. The Fe/Nb@Se$_6$ octahedra and Nb@Se$_6$ prismatic chains along with the Se-Se bonds are highlighted. (Se atoms: green; Fe atoms: pink; Nb atoms: blue.) (*b*) The needle-shaped crystals of FeNb$_3$Se$_{10}$.

**Table 1.** The crystal structure refinement parameters for FeNb$_3$Se$_{10}$ at room temperature at pressures up to 4.4 GPa.

| Pressure | 0 GPa | 2.5 GPa | 4.4 GPa |
|---|---|---|---|
| Space Group | $P2_1/m$ | $P2_1/m$ | $P2_1/m$ |
| Unit cell dimensions | $a = 9.2204(5)$ Å<br>$b = 3.4614(1)$ Å<br>$c = 10.1514(6)$ Å<br>$\beta = 113.846(7)°$ | $a = 9.054(2)$ Å<br>$b = 3.4507(5)$ Å<br>$c = 10.148(3)$ Å<br>$\beta = 114.47(3)°$ | $a = 8.953(3)$ Å<br>$b = 3.4360(5)$ Å<br>$c = 10.097(3)$ Å<br>$\beta = 114.66(4)°$ |
| Volume | 296.33(3) Å$^3$ | 288.6(1) Å$^3$ | 282.3(1) Å$^3$ |
| Total Reflections | 14214 | 3755 | 3683 |
| Independent reflections | 3313 [$R_{int} = 0.0518$] | 821 [$R_{int} = 0.1813$] | 801 [$R_{int} = 0.1890$] |
| Refined Parameters | 45 | 45 | 45 |
| Final R indices (I>2σ(I)) | $R_1 = 0.0356$<br>$wR_2 = 0.0646$ | $R_1 = 0.1594$<br>$wR_2 = 0.3535$ | $R_1 = 0.1554$<br>$wR_2 = 0.3592$ |
| Final R indices (All) | $R_1 = 0.0652$<br>$wR_2 = 0.0721$ | $R_1 = 0.2576$<br>$wR_2 = 0.4315$ | $R_1 = 0.2635$<br>$wR_2 = 0.4553$ |
| Largest Diff. Peak | +5.324 e/Å$^{-3}$ | +7.251 e/Å$^{-3}$ | +7.255 e/Å$^{-3}$ |
| Deepest Diff. Hole | -4.313 e/Å$^{-3}$ | -4.122 e/Å$^{-3}$ | -6.242 e/Å$^{-3}$ |
| R.M.S. | 0.532 | 1.224 | 1.939 |
| Goodness-of-fit on F$^2$ | 1.045 | 1.304 | 1.473 |

**Table 2.** The atomic coordinates and equivalent isotropic atomic displacement parameters for FeNb$_3$Se$_{10}$ at room temperature at pressures up to 4.4 GPa. ($U_{eq}$ is defined as one third of the trace of the orthogonalized $U_{ij}$ tensor.)

| 0 GPa | Wyck. | x | y | z | Occ. | U$_{eq}$ |
|---|---|---|---|---|---|---|
| Nb | 2e | 0.77609(4) | ¼ | 0.86525(4) | 1 | 0.00675(6) |
| Nb/Fe | 2e | 0.05067(7) | ¼ | 0.40756(6) | 0.415(7)/0.585 | 0.0153(1) |
| Se1 | 2e | 0.45883(5) | ¼ | 0.25640(4) | 1 | 0.00870(7) |
| Se2 | 2e | 0.15951(6) | ¼ | 0.66393(6) | 1 | 0.0177(1) |
| Se3 | 2e | 0.33890(5) | ¼ | 0.00498(4) | 1 | 0.00724(6) |
| Se4 | 2e | 0.01164(5) | ¼ | 0.13719(4) | 1 | 0.00545(6) |
| Se5 | 2e | 0.74709(6) | ¼ | 0.58910(4) | 1 | 0.01013(7) |

| 2.5 GPa | Wyck. | x | y | z | Occ. | $U_{eq}$ |
|---|---|---|---|---|---|---|
| Nb | 2e | 0.7701(10) | ¼ | 0.8634(6) | 1 | 0.063(5) |
| Nb/Fe | 2e | 0.0525(15) | ¼ | 0.4086(9) | 0.40(6)/0.60 | 0.064(7) |
| Se1 | 2e | 0.4710(12) | ¼ | 0.2611(7) | 1 | 0.055(5) |
| Se2 | 2e | 0.1638(14) | ¼ | 0.6688(8) | 1 | 0.062(6) |
| Se3 | 2e | 0.3415(12) | ¼ | 0.0056(6) | 1 | 0.062(5) |
| Se4 | 2e | 0.0102(12) | ¼ | 0.1360(6) | 1 | 0.062(5) |
| Se5 | 2e | 0.7373(13) | ¼ | 0.5858(7) | 1 | 0.076(6) |

| 4.4 GPa | Wyck. | x | y | z | Occ. | $U_{eq}$ |
|---|---|---|---|---|---|---|
| Nb | 2e | 0.7669(12) | ¼ | 0.8627(6) | 1 | 0.050(3) |
| Nb/Fe | 2e | 0.0556(17) | ¼ | 0.4088(9) | 0.35(8)/0.65 | 0.046(5) |
| Se1 | 2e | 0.4783(14) | ¼ | 0.2612(7) | 1 | 0.048(4) |
| Se2 | 2e | 0.1617(16) | ¼ | 0.6691(8) | 1 | 0.057(4) |
| Se3 | 2e | 0.3433(13) | ¼ | 0.0035(7) | 1 | 0.042(3) |
| Se4 | 2e | 0.0120(13) | ¼ | 0.1367(6) | 1 | 0.043(3) |
| Se5 | 2e | 0.7365(13) | ¼ | 0.5848(7) | 1 | 0.047(3) |

**Ambient Pressure Insulating FeNb$_3$Se$_{10}$:** The temperature-dependent resistivity data, depicted in **Fig. 2**, shows distinctly different behavior in different temperature ranges. From 110 K to 300 K, the data is consistent with thermal excitation across an energy gap, fitting the equation $e^{\frac{\Delta}{2k_B T}}$, yielding an excitation energy of approximately 21 meV. Between 50 K and 110 K, the excitation energy is approximately 29 meV. In the range of 23 K to 45 K, the resistivity data fits better to $e^{\left(\frac{T_0}{T}\right)^{1/4}}$, suggesting possible Anderson localization contributions to the resistance due to the random potential created in the disordered Fe-Nb octahedral chains.[18] At even lower temperatures, from 1.8 K to 20 K, no simple form is observed, which is likely due to the interference of the measurement system. Previous study suggests a CDW transition ~ 140 K, but the CDW transition is not clear in our data.[40]

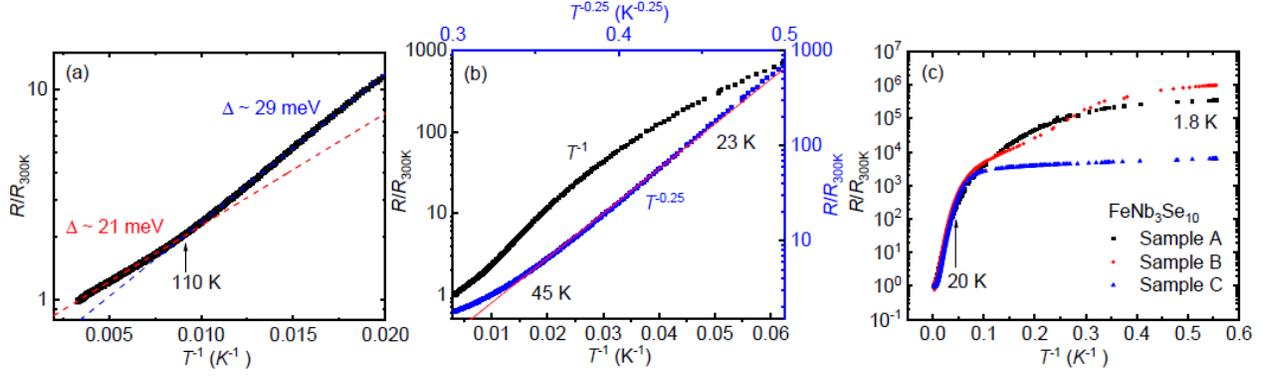

**Fig. 2. Electrical resistivity measurements on FeNb$_3$Se$_{10}$ at ambient pressure.** (*a*) fitting with the $e^{\frac{\Delta}{2k_BT}}$ equation from 110 to 300 K and from 50 K to 110 K; (*b*) fitting with the $e^{\left(\frac{T_0}{T}\right)^{1/4}}$ equation from 23 to 45 K; (*c*) the temperature-dependent resistivity data from 1.8 to 20 K.

To support the experimentally determined electronic properties theoretically, we utilize first-principles calculations to demonstrate that a hypothetical model of FeNb$_3$Se$_{10}$ with an ordered arrangement of Fe and Nb atoms in the octahedra exhibits semimetallic properties when spin-obit coupling (SOC) is not included, as illustrated in **Fig. 3a**. Upon inclusion of spin–orbit coupling for Fe and Nb atoms, a small energy gap ($\approx$ 18 meV) emerges at the line nodes, while the overall bulk band structure retains its semimetallic/insulating characteristics. These findings align with the electrical transport measurements shown in **Fig. 2**. The orbital projections (**Fig. 3a** and **Fig. S1**) indicate that these bands originate from Fe 3*d* and Nb 4*d* orbitals. Along the Γ–B direction, the band crossing is gapless, with both bands around the crossing point sharing the same slope. In contrast, a band gap exists along the Γ–Z direction. In a spinless solid-state system with time-reversal symmetry, it is possible that node points could form a one-dimensional line rather than discrete points along the Γ–B direction, as the spinless Hamiltonian of two bands can always be real-valued, resulting in nodal solutions with a dimension of one in three-dimensional momentum space. With spin–orbit coupling, an energy gap is introduced along the nodal line, potentially transforming the material into a strong topological insulator. The size of this spin–orbit gap is approximately 18 meV. Despite the potential nodal line being gapped in calculations with spin–orbit coupling which requires further theoretical study, the two bulk bands around the gap maintain the same slope, indicating similar low-energy behavior of bulk carriers, as shown in **Fig. 3c**.

To validate this interpretation of the electronic structures of hypothetical models, we synthesized long single crystals (~1 mm ribbon-like) of FeNb$_3$Se$_{10}$ and conducted angle-resolved photoemission spectroscopy (ARPES) measurements. The monoclinic nature of the crystals makes measuring exact high symmetry lines challenging, however, due to the low dispersion in $k_z$, we can still get useful information by using a projected Brillouin zone (**Fig. 3d**). ARPES measurements taken at a photon energy of 138 eV map the electronic structure of FeNb$_3$Se$_{10}$ at ambient pressure along the high symmetry lines of this projected Brillouin zone. The mapping of the iso-energy cut is shown in **Fig. 3e** with the projected high symmetry points and unit cell boundaries indicated with white dashed lines. **Fig. 3f** shows the band structure along the projected $\bar{Z} - \bar{Q}$ line. Here, the gap size is approximately 0.2 eV. Please note that this 0.2 eV gap may not represent the actual gap size of the material, as the gap along the $\bar{\Gamma} - \bar{B}$ direction is much smaller. The ARPES results corroborate first-principal band calculations and electrical transport measurements, revealing the presence of a small gap in the material Unfortunately, we were not able to confirm the gap along the Γ–B direction. This is because the material exhibits van der Waals bonding along the crystallographic *c*-direction, which corresponds to the Γ–B direction, making it impossible to cleave the side surface. This non-metallic property distinguishes FeNb$_3$Se$_{10}$ from non-doped TaSe$_3$ and NbSe$_3$. We do not observe distinct features of the CDW transitions in our data.

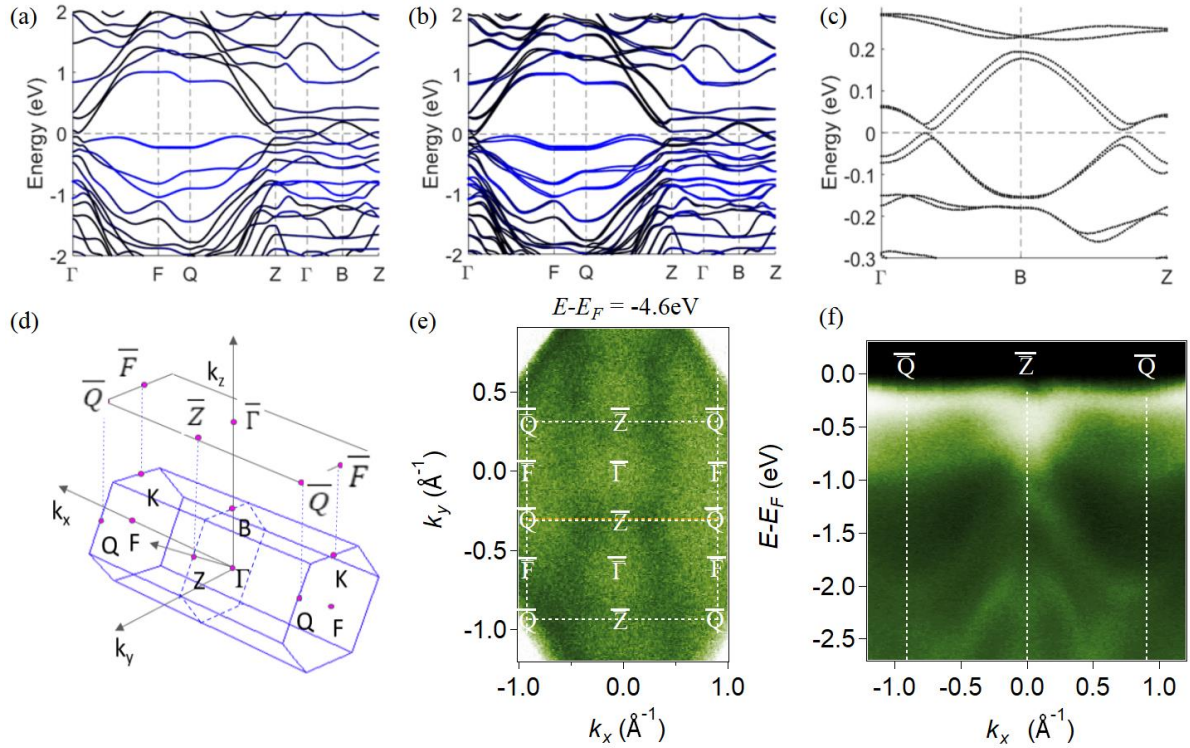

**Fig. 3. Electronic Structure of FeNb$_3$Se$_{10}$.** (*a*) without spin orbit coupling (SOC) (*b*) when spin orbit coupling is included for the Fe and Nb atoms; The blue color indicates the Fe d orbital contribution (*c*) magnified band structure illustrating the band gap around the Fermi level when spin orbit coupling is included; (*d*) A map of the Brillouin zone (BZ) with the high symmetry points marked (a rectangular-shape projected BZ is shown above; ) (*e*) The experimental ARPES data of an iso-energy cut taken at 138eV with the projected BZ marked; (*f*) experimental band cut taken at the orange marked line in (*e*) showing good agreement with calculated bulk bands.

**Insulator to Metal Transition:** High pressure serves as a clean and effective tool for tuning lattice parameters, significantly impacting the electronic structure of materials. To explore the effect of pressure on the transport properties of FeNb$_3$Se$_{10}$, we performed high-pressure resistivity measurements on that material as a function of temperature, *R*(*T*), using a DAC with current applied along the long edge of the ribbon-like crystal (sample #1), consistent with the ambient pressure structural measurements presented in **Fig. 1**. Given the quasi-one-dimensional nature of the crystal structure, we hypothesized that electrical resistance would exhibit significant directional dependence. The results, shown in **Fig. 4a**, reveal that at 1.0 GPa and 2.0 GPa, the resistance displays insulating behavior, with a monotonic decrease in thermal activation energy near room temperature (see **Fig. 4b**). Notably, a distinct insulator-to-metal transition as observed at about 3.0 GPa, where the sample exhibits metallic behavior from 300 K down to ~50 K, with a slight upturn in resistance below 50 K. Upon further increasing the pressure, the sample demonstrates fully metallic behavior in the full temperature range.

By examining the crystal structures of FeNb$_3$Se$_{10}$ obtained at 2.5 GPa, it is noteworthy that the Nb/Fe-Se bond distances within the Nb/Fe@Se$_6$ octahedra, which range from 2.381(1) to 2.620(1) Å at ambient pressure, appear to increase slightly, to 2.405(5) to 2.631(2) Å at 2.5 GPa. Similarly, the Nb-Se bond distances within the Nb@Se$_6$ prismatic polyhedra show a slight increase from 2.6203(5)-2.6551(5) Å at ambient pressure to 2.623(3)-2.656(2) Å at 2.5 GPa. This occurs despite the observation that the unit cell volume decreases with increasing pressure as expected. This indicates that while the intrachain distances have increased in this pressure range, the interchain distances have decreased, which may contribute to the observed transition from insulating to metallic behavior.

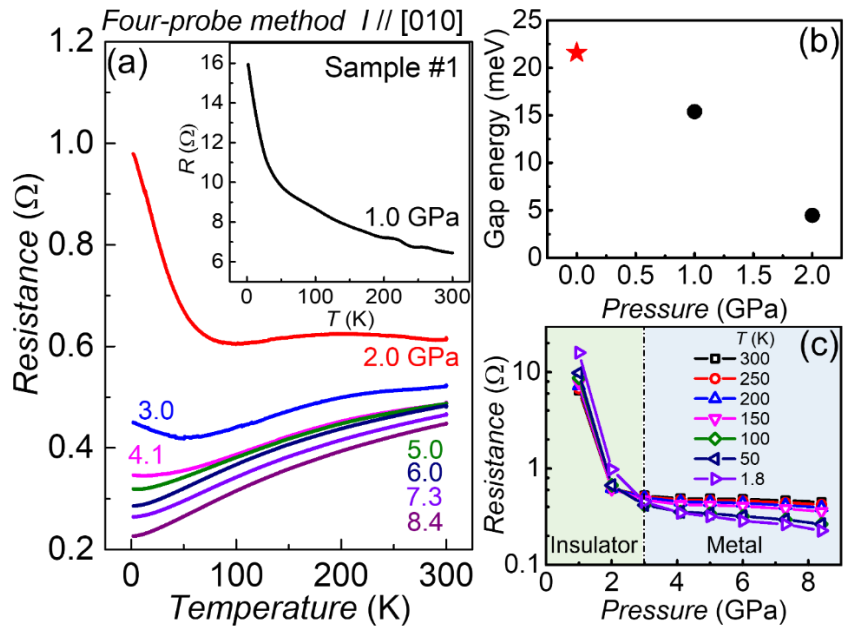

**Fig. 4.** (*a*) Pressure-dependent resistance of FeNb$_3$Se$_{10}$ single crystal sample #1 up to 8.4 GPa and down to 1.8 K. The current direction is along the long side of ribbon-like crystal. The inset shows *R(T)* curve measured at 1 GPa. (*b*) The estimated resistivity gap near room temperature for the crystal at ambient pressure (red star), estimated from the data in Fig. 2, and at 1.0 GPa and 2.0 GPa. (*c*) Comparison of pressure dependent resistance at various temperatures from 1.8K to 300 K. Note that the y-axis is a log scale.

**The "Bad Metal" Behavior at Higher Pressure:** The observation of the insulator-to-metal transition above ~3 GPa prompts us to further investigate the potential emergence of superconductivity above 1.8 K under pressure, analogous to the superconductivity observed in $NbSe_3$[41] at ambient pressure. **Fig. 5a** presents the $R(T)$ measurements of sample #2. Different from the measurement on sample #1, the $R(T)$ of sample #2 was measured up to ~ 66 GPa under quasi-hydrostatic pressure using the van der Pauw method instead of the four-probe method. In contrast to the measurements on sample #1, this approach may average the resistivity across different crystallographic directions. At the maximum pressure of the measurements, no visible deformation or cracks were observed in the sample, and the Nujol oil, employed as the pressure-transmitting medium, remained transparent throughout the entire pressure range.

Metallic behavior was observed in sample #1 between about 3.0 GPa and 8.5 GPa. This leads us to compare the crystal structure and atomic coordinates in $FeNb_3Se_{10}$ at 2.5 and 4.4 GPa. In contrast to the clear trend observed in atomic distances from 0 to 2.5 GPa, the Nb/Fe-Se bond distances within the disordered Nb/Fe@$Se_6$ octahedra, which range from 2.405(1) to 2.631(1) Å at 2.5 GPa, decrease slightly, to 2.393(5) and 2.609(2) Å at 4.4 GPa. Similarly, the Nb-Se bond distances within the Nb@$Se_6$ prismatic polyhedra decreased only slightly, from 2.623(3)-2.656(2) Å at 2.5 GPa to 2.619(3)-2.654(2) Å at 4.4 GPa, despite the continued reduction in unit cell volume with increasing pressure. These observations suggest that the interchain distances continue to decrease with pressure.

Clear metallic behavior in sample #2 was detected only below ~100 K in the pressure range of 3.5 GPa to 10.7 GPa, and below ~50 K in the range of 19.6 GPa to 30.2 GPa, as shown in **Fig. 5a**. The minor discrepancy can potentially be attributed to two factors: first, the sensitivity of the measurements to the hydrostaticity of the pressure (stress) conditions; and second, the intrinsic anisotropy of quasi-1D materials, where electronic conduction is typically highest along specific crystallographic directions (such as the chain direction in the current material). Interestingly, as pressure increases further, the overall $R(T)$ curve exhibits a sudden drop at about 32 GPa and becomes insensitive to additional pressure increases (see **Fig. 5b**). This behavior may suggest that higher pressure alters the crystal structure profoundly, potentially increasing the effective mass of the charge carriers or reducing the band overlap. The abrupt decrease in overall resistance beyond

35 GPa likely indicates a lattice or electronic structural transition, warranting further investigation to fully understand this unusual phenomenon.

Notably, no superconducting transition was detected above 1.8 K in any of the high-pressure measurements on FeNb$_3$Se$_{10}$, unlike what is observed for NbSe$_3$. However, additional measurements at temperatures below 1.8 K and higher pressures are required to definitively confirm the absence of superconductivity in the metallic state of this compound.

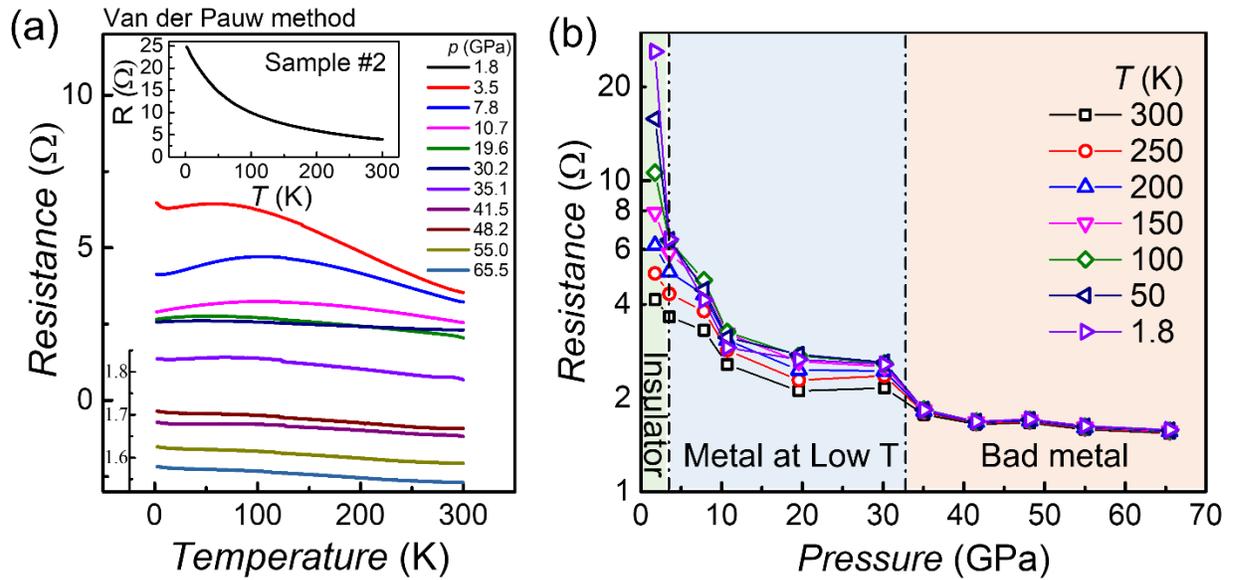

**Fig. 5.** (*a*) The temperature and pressure-dependent resistance of a FeNb$_3$Se$_{10}$ single crystal, sample #2, up to 65.5 GPa and down to 1.8 K. The Van der Pauw method was used to measure the resistance. The upper inset shows the *R*(*T*) curve measured at 1.8 GPa, while the lower inset presents the *R*(*T*) curves from 35.1 GPa to 65.0 GPa, with the resistance values plotted on a smaller scale to provide better clarity of the curves' behaviors. (*b*) Comparison of the pressure dependent resistance at various temperatures from 1.8K to 300 K. Note that the y-axis is a log scale.

## Conclusions

The pressure-dependent crystal structure at room temperature, as well as the temperature and pressure dependencies of the resistivity, have been systematically investigated for FeNb$_3$Se$_{10}$. Structural refinement is consistent with studies that indicate that the formula is Fe-rich of FeNb$_3$Se$_{10}$, closer to Fe$_{1.20}$Nb$_{2.80}$Se$_{10}$, consistent with some previous studies. Our theoretical studies find that in the absence of spin orbit coupling the material is a metal, but that the inclusion of SOC on the Nb and Fe atoms opens up a very small gap at the Fermi energy suggesting that the material with formula $FeNb_3Se_{10}$ may be topological semimetal, though further proof is needed to determine whether that is actually the case. The material is shown to undergo an insulator to metal transition at the modest pressure of about 3.5 GPa, while maintaining its crystal structure, but unlike $NbSe_3$, superconductivity is not observed.

## Acknowledgements

The work at Michigan State University was supported by the U.S. Department of Energy (DOE), Division of Basic Energy Sciences under Contract DE-SC0023648. The work performed at Ames National Laboratory (SH, SLB, PCC) was supported by the U.S. Department of Energy, Office of Basic Energy Science, Division of Materials Sciences and Engineering. Ames National Laboratory is operated for the U.S. Department of Energy by Iowa State University under Contract No. DE-AC02-07CH11358. The work at Princeton University was funded by the US Department of Energy, grant DE-FG02-98ER45796. The work at the University of Michigan was supported by the NSF CAREER grant under Award No. DMR-2337535. S.S. was funded by the National Science Foundation (NSF) through the Materials Research Science and Engineering Center at the University of Michigan, Award No. DMR-2309029. This work used resources of the Advanced Light Source, a U.S. Department of Energy (DOE) Office of Science User Facility under Contract No. DE-AC02-05CH11231. This work used resources of the Advanced Light Source, a U.S. Department of Energy (DOE) Office of Science User Facility under Contract No. DE-AC02-05CH11231.

# Table of Contents

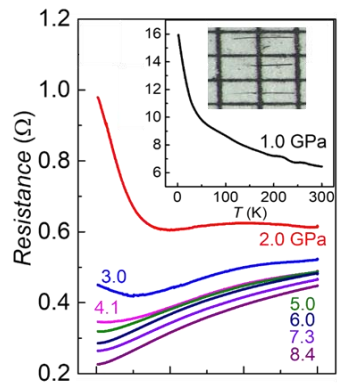